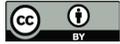

Earth System Dynamics

# Hydrological cycle over South and Southeast Asian river basins as simulated by PCMDI/CMIP3 experiments

S. Hasson[1,2], V. Lucarini[1,3], and S. Pascale[1]

[1]Meteorological Institute, KlimaCampus, University of Hamburg, Hamburg, Germany
[2]Institute of Geography, University of Hamburg, Hamburg, Germany
[3]Department of Mathematics and Statistics, University of Reading, Reading, UK

*Correspondence to:* S. Hasson (shabeh.hasson@zmaw.de)

**Abstract.** We investigate how the climate models contributing to the PCMDI/CMIP3 dataset describe the hydrological cycle over four major South and Southeast Asian river basins (Indus, Ganges, Brahmaputra and Mekong) for the 20th, 21st (13 models) and 22nd (10 models) centuries. For the 20th century, some models do not seem to conserve water at the river basin scale up to a good degree of approximation. The simulated precipitation minus evaporation ($P - E$), total runoff ($R$) and precipitation ($P$) quantities are neither consistent with the observations nor among the models themselves. Most of the models underestimate $P - E$ for all four river basins, which is mainly associated with the underestimation of precipitation. This is in agreement with the recent results on the biases of the representation of monsoonal dynamics by GCMs. Overall, a modest inter-model agreement is found only for the evaporation and inter-annual variability of $P - E$. For the 21st and 22nd centuries, models agree on the negative (positive) changes of $P - E$ for the Indus basin (Ganges, Brahmaputra and Mekong basins). Most of the models foresee an increase in the inter-annual variability of $P - E$ for the Ganges and Mekong basins, thus suggesting an increase in large low-frequency dry/wet events. Instead, no considerable future change in the inter-annual variability of $P - E$ is found for the Indus and Brahmaputra basins.

## 1 Introduction

One of the most relevant and debated aspects of climate change deal with its impacts on the hydrological cycle, and, in particular, on statistical properties and seasonality of the precipitation, evaporation, runoff, and, consequently, on the discharge of the rivers. At present, the agriculture-based economies of South and Southeast Asia face serious challenges in ensuring the food security and economic well-being of 1.4 billion people due to their large dependence on the variable freshwater supplies and inadequate integrated water resources management (Lal et al., 2011). South and Southeast Asia are hot spots of climate change and it is anticipated that the changes in the hydrological cycle will be quite serious in this region, providing further challenges to the existing water management problems, with potentially serious impacts on its socio-economical processes (IPCC, 2007). Therefore, in terms of public policy, the water resource management and the defense from the hydrological risks cannot be decoupled from the climate change agenda (Ruelland et al., 2012). As an example, considering the future water availability under the warmer climate while planning new water reservoirs can be long-sighted, keeping in mind the huge investment costs and the fact that the lifetime of these reservoirs is comparable to the timescale over which significant changes in the hydro-climatology of the region will manifest themselves clearly (Krol and Bronstert, 2007). Hence, it is crucial for the policy makers and the regional actors to have high-quality information on the projected climate change and its implications for the water resources of the region.

Atmosphere-ocean coupled general circulation models (GCMs) are the most powerful tools currently being used by the scientific community for studying climate change and its impacts on the hydrological cycle (Fowler et al., 2007; Johnson and Sharma, 2009). However, a realistic representation of the hydrological cycle at both global and regional scale is indeed a complex task (Lucarini et al., 2008; Liepert and Previdi, 2012), as non-trivial dynamics occur on a vast range of spatial and temporal scales, often smaller than those explicitly resolved by these models. Furthermore, GCMs feature problems of self-consistency in terms of conservation of water mass on global and regional scales which imply inconsistencies in their energetics (Liepert and Previdi, 2012; Lucarini and Ragone, 2011). Therefore, it seems crucial to investigate the relevant processes in these models before their use in further analysis/applications. In this regard, we investigate whether the water balance of the PCMDI/CMIP3 (Third Phase of the Climate Models Inter-comparison Phase of the Program for Climate Model Diagnosis and Inter-comparison, see http://www-pcmdi.llnl.gov/; Meehl et al., 2004) climate models is closed for the major river basins of South and Southeast Asia (Indus, Ganges Brahmaputra and Mekong), pointing out inconsistencies of the individual models in detail, if any. CMIP3 models have been chosen for two reasons. First, we wish to define a benchmark for evaluating the performance of the new generation of GCMs (i.e. CMIP5), which have undergone extensive development and modifications, introducing higher resolutions, atmosphere and land use and vegetation interaction, detailed aerosols treatment, carbon cycle, etc. (Taylor et al., 2012). Secondly, our focus on the CMIP3 is of direct relevance for most of the South and Southeast Asian downscaling communities and institutions involved in operational activities – e.g. the Pakistan Meteorological Department and the Global Change Impact Studies Centre (Pakistan), the Department of Hydrology and Meteorology (Nepal), the Indian Institute for Tropical Meteorology (India) and the National Climate Center of Thailand (Thailand) – which are currently using regional climate models nested into CMIP3 models.

As mentioned above, the regional water balance inconsistencies of the climate models may be one of the major causes of biases in their simulated regional-scale hydrological cycle. Provided that the water balance of the CMIP3 models is closed, it is relevant to examine their skill in representing the hydrological cycle over the region to get reliable estimates of their projected changes under the climate change scenarios. Although, several studies have been performed in this direction (Kripalani et al., 1997; Kang et al., 2002; Annamalai et al., 2007; Lin et al., 2008), such studies mainly focused on the representation of the monsoon and its interaction with the westerly disturbances over the region. However, in order to adequately quantify the biases in the hydrological cycle, it is worth investigating its other aspects in addition to precipitation. Moreover, most studies (Arnell, 1999; Nijssen et al., 2001; Douville et al., 2002; Manabe et al., 2004; Hagemann et al., 2006; Nohara et al., 2006) presenting the observed/simulated water balance information are either on a global scale – thus not providing catchment-scale characteristics of the hydrological cycle – or always miss one or the other major South and Southeast Asian river basins.

Following Lucarini et al. (2008), we believe that the verification and validation of GCMs should specifically be performed at a river-basin scale, through accurately calculating the most relevant hydrological quantities within the basin boundaries, in order to characterize the models' behavior within the naturally defined geographical units relevant for the water resource management. This allows for conducting reliable impact assessment studies, and subsequently their applicability for the informed decisions by the bodies responsible for the water resources management in the region. Thus, our present study tries to fill a gap by addressing the basin scale representation of the hydrological cycle by the individual/cluster of models, for all the major river basins of South and Southeast Asia region.

Large inter-model variability is a major problem in ascertaining the reliability of future climate change projections. To summarize the output of multi-model datasets, a common approach is to take the arithmetic mean of the ensemble members and the standard deviation as a measure of spread (Houghton et al., 2001; Milly et al., 2005; Nohara et al., 2006). The approach, originally developed for seasonal forecasting (Harrison et al., 1999; Fritsch et al., 2000), has been extended to multi-model climate projections by weighting each GCM according to its performance for the present climate (Giorgi and Mearns, 2002). However, considering the output of each considered climate model as an ensemble member is indeed problematic, for multiple reasons discussed in detail by Lucarini et al. (2008) and Liepert and Previdi (2012) and briefly summarized here: (i) there is no way to correctly assign weights to climate models in terms of their quality due to huge inter-model structural differences; (ii) climate models' output do not form a sample from any well-defined probability distribution. Moreover, GCMs feature systematic spatio-temporal biases. Therefore, taking the ensemble mean as the representative output from the multi-model simulations does not really take care of removing the inconsistencies or inaccuracies present in various GCMs, as no actual compensating effect coming from errors of different sign can be really invoked. These ensemble-mean based estimates can therefore be misleading for the impact assessment studies and consequently their use in spatial planning is questionable. In order to have a semi-quantitative evaluation of how well the ensemble mean is representative of the whole dataset of GCMs outputs, we introduce a simple indicator measuring the ratio between the spread of the models' outputs – defined as the difference between the largest and smallest value given by the various GCMs – and their arithmetic average.

The annual mean and the deviations from the long-term mean are calculated for hydrological observables such as the basin-integrated precipitation ($P$), evaporation ($E$), $P - E$ and runoff ($R$). We first considered the last 40 yr of the 20th century in order to verify whether the land modules of the climate models conserve water, and in order to assess the realism by comparing the simulated quantities with historical

Earth Syst. Dynam., 4, 199–217, 2013 www.earth-syst-dynam.net/4/199/2013/

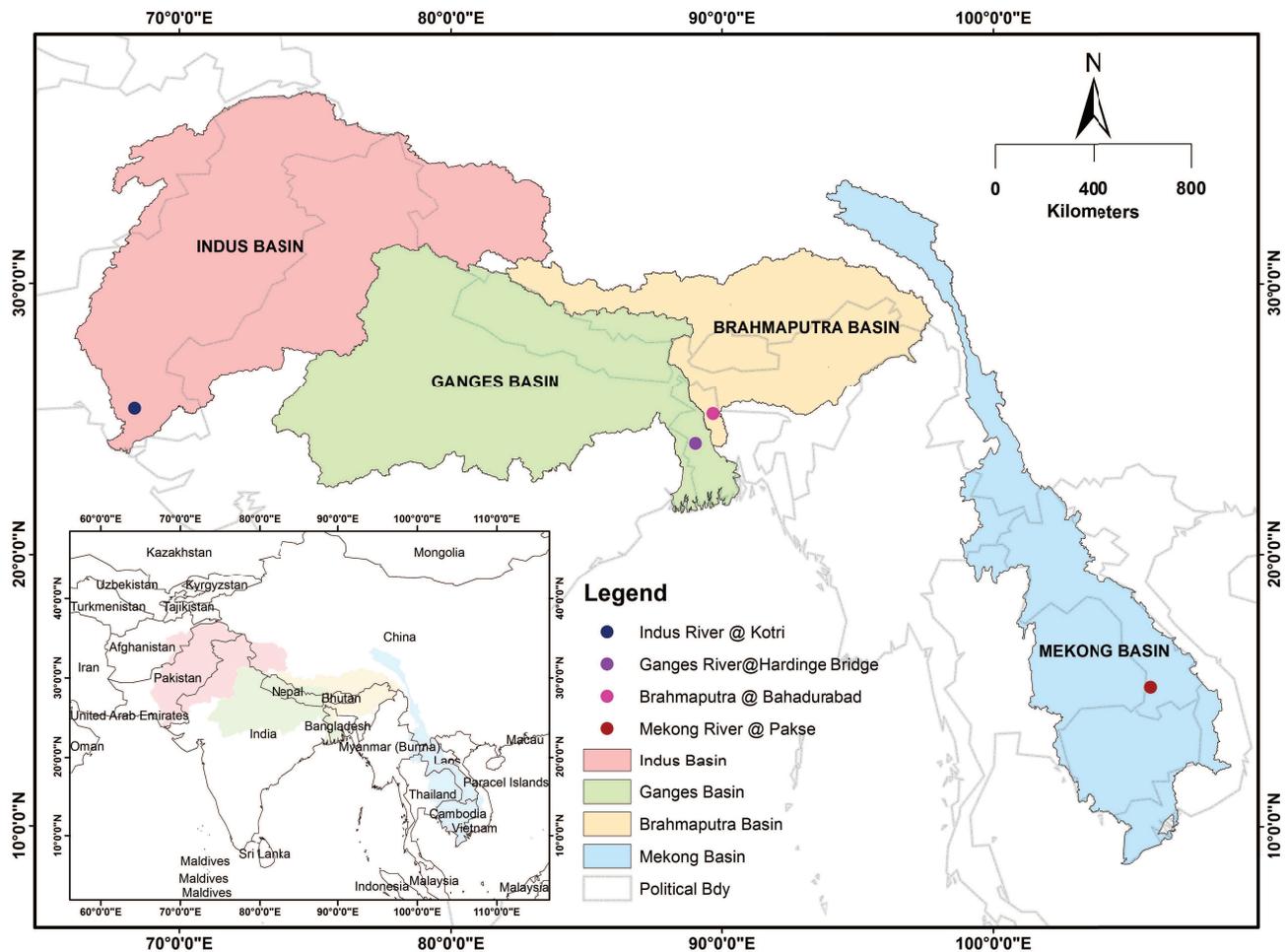

**Fig. 1.** Four study river basins: Indus, Ganges, Brahmaputra and Mekong (west to east).

observations of precipitation and river discharges. Moreover, future changes for the same hydrological quantities for the last 40 yr of the 21st and of the 22nd centuries with respect to the 20th century (1961–2000) are assessed by considering the SRES A1B scenario. The SRESA1B scenario (720 ppm of $CO_2$ after 2100) is chosen, as it represents the median of the rest of the IPCC scenarios in terms of the greenhouse gas forcing (GHG) (IPCC, 2007).

## 2 Study region

The study region comprises four major river basins of South and Southeast Asia namely the Indus, Ganges, Brahmaputra and Mekong (see Fig. 1). The hydrological regimes of these rivers differ because their basins feature a great geographical and climatic diversity, including latitudinal and longitudinal extent, regional climate patterns, influence of nearby seas and desert, positions with respect to the Hindu Kush–Himalayan (HKH) range and presence or lack of extensive cryosphere. As a common ground, the hydrology of these basins depends on the moisture input from the monsoon system, with snow and glacier melt (rainfall) being extremely relevant at high (low) altitudes, particularly for the Indus and Brahmaputra basins. The eastern basins of the study area are comparatively wetter than the western basins because the monsoon rainfall dominates in the summer months in the eastern part and gets weaker on the western side with a time delay of some weeks. In the west, westerly disturbances drop moisture in the winter months mainly in the form of solid precipitation (Rees and Collins, 2006). This effect is much weaker as we go east, because the Atlantic–Mediterranean storm track does not extend beyond the HKH (Bengtsson et al., 2007). This fact is particularly evident from the significance of meltwater contribution, which is reported to be extremely important for the Indus, important for the Brahmaputra, modest for the Ganges and negligible for the Mekong basin (Immerzeel et al., 2010). Table 1 summarizes the general characteristics of the four basins, whereas details of individual basins are discussed below.

Table 1. Characteristics of the four studied river basins.

| Basin characteristics | Indus | Ganges | Brahmaputra | Mekong |
|---|---|---|---|---|
| Basin area (km$^2$) | 1 230 000 | 1 000 000 | 530 000 | 840 000 |
| River length (km) | 3200 | 2500 | 2900 | 4800 |
| Precipitation (mm yr$^{-1}$) | 415 | 1125 | 1350 | 1550 |
| Near-to-sea gauge used in study | Kotri | Hardinge Bridge | Bahadurabad | Pakse |
| Annual mean discharge (m$^3$ s$^{-1}$)/ Runoff (mm yr$^{-1}$) to sea | 1540/ 40 obs. 210 natural | 11 000/ 345 | 20 000/ 1200 | 17 000/ 650 |
| Discharge (m$^3$ s$^{-1}$) equivalence to the 100 mm yr$^{-1}$ runoff | 3890 | 3190 | 1680 | 2670 |
| Peak discharge month | August | August | July | August |
| High flow season | April–September | July–October | April–November | June–November |
| No of glaciers/area (km$^2$) | 18 495/21 000 | 7963/9000 | 11 497/14 000 | 482/230 |
| Snow coverage (Annual Avg. %) | 13.5 | 5 | 20 | 3 |
| Snow and glacier melt index | 150 | 10 | 27 | Negligible |
| Population dependent (millions) | 260 | 520 | 66 | 79 |
| Major consumption | Agriculture | Agriculture | Agriculture | Agriculture |
| Seasonal/annual Variability | High | High | High | High |

## 2.1 Indus Basin

The Indus originates from the Tibetan Plateau (China) and it drains through India, Afghanistan and Pakistan before its confluence to the Arabian Sea. The basin is divided into an upper and a lower part, at the point where the Indus River enters into the Tarbela reservoir in Pakistan. As discussed above, the basin hydrology is dominated by two main moisture sources, i.e. monsoon and western disturbances. The northwestern part of the basin – hosting maximum freshwater resources – is characterized by an increase in the observed precipitations (Archer and Fowler, 2004) and cooling temperature trends (Fowler and Archer, 2005) for the last few decades of the 20th century. However, ensembles of 13 and 17 CMIP3 GCMs project higher warming rates over this part than over the southern plains and against the respective global averages for SRESA2 and A1B scenarios throughout the 21st century (Islam et al., 2009). The average precipitation over the basin is estimated to be about 415 mm yr$^{-1}$ (CRU, 2012). The precipitations are confined to the monsoon (July–September) and winter (December–March) seasons. Almost 80 % of the mean annual flows are confined to the summer months (April–September) with a peak in the month of August, due to snow and glacier melt as well as the monsoonal rainfall. During the winter (October–March), the river experiences low flow conditions. The average discharge into sea, measured at Kotri which is the last gauging site near the sea, is calculated to be around 40 mm yr$^{-1}$ runoff equivalent for the period 1977–2000. Indus River is strongly anthropogenically influenced as roughly 80 % of its total surface water (i.e. 210 mm yr$^{-1}$ or 258 km$^3$ yr$^{-1}$) is diverted for irrigation and other purposes on annual basis (Laghari et al., 2012). The agriculture sector consumes more than 95 % of such amount to ensure the food security and economic well-being of about 260 million people (CIESIN, 2005).

With 18 495 glaciers covering an area of 21 000 km$^2$ (Bajracharya and Shrestha, 2011) and 13.5 % of average snow coverage (Gurung et al., 2011), the Indus River has the highest meltwater index as compared to other South and Southeast Asian rivers originating from the HKH region. Based on a modeling study, Immerzeel et al. (2010) report that the normalized meltwater index for the Indus is 151 % of the total discharge naturally generated downstream. The index is calculated as a ratio between the volume of the upstream snow and glacier melt runoff and $P - E$ naturally available downstream.

## 2.2 Ganges Basin

Originating from the Central Himalayan Range, the Ganges drains across Bangladesh, China, India and Nepal before its confluence to the Indian Ocean at the Bay of Bengal. The river is highly regulated with dams and irrigation canals right after it enters into the plains near Haridwar (Bharati et al., 2011), usually for agriculture purposes. Agriculture has a share of almost 90 % of the available water to ensure the well-being of about 520 million people (CIESIN, 2005). The hydrological regimes of the basin are dominated by the summer monsoon system. Almost 75 % of the annual precipitation occurs during the monsoon season (June–September). The mean annual precipitation estimated from the observed dataset CRU TS3.2 (CRU, 2012) is around 1125 mm yr$^{-1}$. The observational record indicates that the precipitation within the basin is by and large stable (Mirza et al., 1998; Immerzeel, 2008). The mean flow for the period 1950–2008 is estimated as about 11 000 m$^3$ s$^{-1}$ (345 mm yr$^{-1}$) at the Hardinge Bridge, the last gauging site near the sea. These flows show large inter-annual and intra-annual variability, featuring a decreasing trend which may further be attributed to the decreasing strength of the Indian monsoon (Webster et al., 1998; Jian et al., 2009) and/or some other factors. More than 85 % of the river flows are confined to the high flow period (July–October) with the mean maximum during late August due to the heavy contributions from the monsoonal rainfall and glacier melt. During the lean flow period (November–June), contributions mainly come from the snowmelt (April–June), winter rainfall and base flow. Glaciers number 7963 in total, which cover an area of about 9000 km$^2$ (Bajracharya and Shrestha, 2011) whereas the annual average snow cover is about 5 % only (Gurung et al., 2011). A modeling study suggests a low meltwater index of 10 % for the basin (Immerzeel et al., 2010).

### 2.3 Brahmaputra Basin

The Brahmaputra originates from the southwestern part of the Tibetan Plateau and drains through China, India, Bangladesh and Bhutan. The river is a lifeline for approximately 66 million people (CIESIN, 2005). The basin hydrology is dominated by the influence of the monsoon system, whereas some contribution is also received from the western disturbances (roughly 10 %). The mean annual precipitation, estimated from the observed dataset CRU TS3.2 (CRU, 2012), is around 1350 mm yr$^{-1}$. The mean flow for the period 1956–2008 is calculated to be about 20 000 m$^3$ s$^{-1}$ (1200 mm yr$^{-1}$) at Bahadurabad, the last gauging site before its confluence to the Ganges River. The analysis of the observed discharge shows that more than 90 % of the flows are confined to the high flow period (April–early November) with a mean maximum during mid-July. The early rise of the hydrograph in the month of April may be attributed to the snow melt contribution reducing during the late spring or early summer, which is then compensated by the glacier melt. The monsoon rainfall starts couple of weeks earlier as compared to the Ganges basin (Webster et al., 1998; Jian et al., 2009), spanning from late May to September. The lean flow period comprises the winter months only. In addition to 11 497 glaciers covering an estimated area of about 14 000 km$^2$ (Bajracharya and Shrestha, 2011), the basin has 20 % of the annual average snow coverage (Gurung et al., 2011). The Brahmaputra basin has a high meltwater index of 27 %, second in the region after the Indus basin (Immerzeel et al., 2010).

### 2.4 Mekong Basin

Draining through China, Myanmar, Lao PDR, Thailand, Cambodia and Viet Nam, the Mekong is amongst the longest rivers in the world. The hydrological regime of the Mekong Basin primarily depends on the climatic conditions of the alternating wet and dry seasons. The climate is governed by the northeasterly and the southwesterly monsoonal winds. The mean annual precipitation ranges between 1000–1600 mm over the dry region of northeast Thailand to 2000–3000 mm over the wet regions of northern and eastern highlands, which accounts for an average of almost 1550 mm yr$^{-1}$ as estimated from the observed dataset CRU TS3.2 (CRU, 2012). The observed precipitation record reveals that the narrow gorge of the northern Mekong basin receives the smallest amount of precipitation as compared to the remaining part. Almost 90 % of the annual rainfall is received under the southwest monsoon system between May and October (MRC, 2005; FAO, 2008), whereas the dry season spans from November to April. The mean discharge is about 17 000 m$^3$ s$^{-1}$ (650 mm yr$^{-1}$) at Pakse gauging site near the sea. Almost 90 % of the annual discharge takes place during the high flow season (June–November) with a peak in the month of August, while the remaining 10 % during the low flow period (December–May). With only 482 glaciers covering an area of about 230 km$^2$ (Bajracharya and Shrestha, 2011) and only about 3 % of the average annual snow cover (Gurung et al., 2011), meltwater has a negligible contribution to the Mekong River. Eastham et al. (2008) suggest that for an extreme climate change scenario where all the glaciers vanish by 2030, the Mekong River would see an increase of only 80 m$^3$ s$^{-1}$ to its average discharge of 3500 m$^3$ s$^{-1}$ at the Chiang Saen site. The agricultural sector is the major consumer of the freshwater (Johnston et al., 2010), being responsible for the food production, livelihood and economic well-being of almost 79 million people (CIESIN, 2005).

## 3 Data and method

### 3.1 Datasets

Our analysis is based on the last 40 yr data of the 20th century climate reconstructions and of the 21st and 22nd centuries' climate projections based on the SRES A1B scenario runs of the AOGCMs (see Table 2 for details) included in the PCMDI/CMIP3 project (http://www-pcmdi.llnl.gov/). The SRESA1B scenario corresponds to a ramp up of CO$_2$ concentration up to 720 ppm in 2100, with an immediate stabilization of CO$_2$ concentration afterwards. It is widely considered as a reasonable, intermediate scenario for climate projections. We have extended our analysis to the projections of 22nd century climate for two main reasons. First, we wish to analyze the effects of 'committed warming' on the hydrological cycle on a longer timescale, when parts of the transient effects of increasing $CO_2$ vanish. Second, we believe that taking into account a longer time perspective is useful for the long-sighted planning and adaptation policies in the region. We have considered monthly values of the relevant climatic variables such as the total runoff ($R$), precipitation ($P$) and evaporation ($E$). The evaporative fields have been reconstructed from the surface upward latent heat fluxes. Our investigations are restricted to 13 climate models for the 20th and 21st century climates, whereas only 10 models provide the complete datasets for $R$, $P$, and $E$ quantities for 22nd century climate. Moreover, in order to understand the degree of realism of climate models in the reconstruction of the 20th century climate, we have considered the observed precipitation and river discharges for each basin. Regarding the observed precipitation, we consider the University of East Anglia Climatic Research Unit (CRU) Time Series (TS) high resolution gridded data version 3.2 (CRU, 2012). However, it is pertinent to mention here that due to the uneven spatial and temporal coverage of the used gauging stations, the presence of extremely non-homogenous terrain, the need for applying data quality control methods and the unavoidable uncertainties coming from the applied interpolation techniques, such gridded rainfall dataset may feature uncertainties which are hard to assess (Fekete et al., 2004; Yatagai et al., 2012). Therefore, caution is needed when interpreting the results based on such gridded datasets. We also consider the historical observed discharges ($D$) for all the rivers at either the last or near-to-sea gauging stations (Indus at Kotri, Ganges at Hardinge Bridge, Brahmaputra at Bahadurabad, and Mekong at Pakse) depending on their data availability. The discharge data for the Indus is collected from the Water and Power Development Authority (WAPDA), Pakistan, for the Ganges and Brahmaputra from Jian et al. (2009), and for the Mekong River from Dai and Trenberth (2002).

Table 2. List of GCMs used in the study. These constitute the subset of all GCMs included in the PCMID/CMIP3 project providing all the climate variables of our interest.

| Name and reference | Institution | Grid resolution (Lat × Lon) |
|---|---|---|
| MRI-CGCM2.3.2 Yukimoto and Noda (2002) | Meteorological Research Institute, Japan Meteorological Agency, Japan | T42 |
| CNRMCM3.0 Salas-Mélia et al. (2005) | Météo-France/Centre National de Recherches Météorologiques, France | T63 |
| CSIRO3.0 Gordon et al. (2002) | CSIRO Atmospheric Research, Australia | T63 |
| ECHAM5 Jungclaus et al. (2006) | Max Planck Institute for Meteorology, Germany | T63 |
| ECHO-G Min et al. (2005) | MIUB, METRI, and M & D, Germany/Korea | T30 |
| GFDL2.0 Delworth et al. (2005) | US Dept. of Commerce/NOAA/Geophysical Fluid Dynamics Laboratory, USA | 2.5° × 2.0° |
| UKMOHADCM3 Johns et al. (2003) | Hadley Centre for Climate Prediction and Research/Met Office, UK | 2.75° × 3.75° |
| UKMOHADGEM1 Johns et al. (2006) | Hadley Centre for Climate Prediction and Research/Met Office, UK | 1.25° × 1.875° |
| INMCM3.0 Volodin and Diansky (2004) | Institute for Numerical Mathematics, Russia | 5° × 4° |
| IPSL-CM4 Marti et al. (2005) | Institute Pierre Simon Laplace, France | 2.4° × 3.75° |
| MIROC (hires) K-1 Model Developers (2004) | CCSR/NIES/FRCGC, Japan | T106 |
| PCM1MODEL Meehl et al. (2004) | National Centre for Atmospheric Research, USA | T42 |
| GISS-AOM Lucarini and Russell (2002) | NASA/Goddard Institute for Space Studies, USA | 4° × 3° |

### 3.2 Theoretical framework

Assuming that the water storage in the liquid and solid form is negligible in the column comprising a soil layer and atmosphere aloft when the long-term averages are considered (Peixoto and Oort, 1992; Karim and Veizer, 2002), at any point of the land surface, the fields $P$, $E$ and $R$ satisfy the balance equation

$$\langle P \rangle_t - \langle E \rangle_t \approx \langle R \rangle_t \approx -\langle \nabla_H \cdot \mathbf{Q} \rangle_t, \quad (1)$$

where the subscript $t$ denotes the temporal averages, and $\nabla_H \cdot \mathbf{Q}$ is the divergence of water in the atmospheric column. Spatially integrating Eq. (1) over an area $A$ of the river basin, the hydrological balance can be written as follows:

$$\int_A dx\, dy\, (\langle P \rangle_t - \langle E \rangle_t) \approx -\int_A dx\, dy\, \langle \nabla_H \cdot \mathbf{Q} \rangle_t$$
$$\approx \int_A dx\, dy\, \langle R \rangle_t \approx \langle D \rangle_t, \quad (2)$$

where $D$ is the observed discharge into the sea. Further details regarding the method and its suitability are discussed in detail by Lucarini et al. (2008). Equation (2) is satisfied for the short-term storages, as the average time of water in the atmosphere is roughly 10 days, whereas routing time in the channels, snow accumulation and the groundwater storages range from month(s) to a season. Equation (2) also provides an excellent approximation for describing the hydrological balance of a river basin if $t \geq 1$ yr, assuming negligible changes in the inter-annual mass balance of existing glaciers, or if the glaciers' mass balance gives minor corrections to the overall hydrological cycle.

An overall small change is observed in the last decade for the mass balance of Karakoram glaciers (Scherler et al., 2011), which is relevant for the Indus basin. Though a negative trend in the glaciers' mass is reported in the central and eastern part of the HKH, the overall correction due to this effect is rather small for the Ganges, Brahmaputra and Mekong basins due to their weak dependence on snow and ice melt contribution as discussed earlier. Therefore, Eq. (2) is appropriate for our case studies.

Yearly time series of the spatially integrated four variables ($P$, $E$, $P - E$ and $R$) are computed for the considered time spans of 40 yr:

$$\overline{\beta}_j = \int_A dx\, dy\, \langle \beta_j \rangle_t, \quad (3)$$

where $\overline{\beta}_j$, $j = 1, \ldots, 4$ corresponds to each of the four variables mentioned above, and $A$ denotes the area of each considered river basin. The long-term averages and standard deviation of the yearly values are computed using Eqs. (4) and (5).

$$\mu(\overline{\beta}_j) = \frac{1}{40} \sum_{i=1}^{40} \overline{\beta}_j \quad (4)$$

$$\sigma(\overline{\beta}_j) = \sqrt{\left[\frac{1}{39} \sum_{i=1}^{40} (\overline{\beta}_j - \mu(\overline{\beta}_j))^2\right]}. \quad (5)$$

### 3.3 Data manipulation

The mean annual time series of all the relevant variables defined in the gridded domain of the climate models are computed using Voronoi or Thiessen tessellation method (Okabe et al., 2000; Lucarini et al., 2008) in the GrADS (Grid Analysis and Display System) and GIS environment. The Thiessen tessellation method has been used to avoid any kind of interpolation, which may prevent the accurate computation of the volumetric quantities, usually along the perimeter of the study basin. The output has been projected to UTM (Universal Transverse Mercator) projection according to the central zone of each river basin, covering the relative maximum of the basin area. For the grid cells partially lying inside/outside the basin, only the fraction of area inside the basin has been considered to prevent water loss/gain. In this way, basin-wide integrated quantities are computed accurately. Similarly, inconsistencies between the land-sea mask given by GCMs and the basin boundaries exist as a result of the fact that each GCM has its own resolution whereas the basin boundary is mapped as the total area accumulating water to a common outlet. Differences are usually relevant for the coarse resolution GCM datasets. So any grid cell partially lying inside/outside the basin at the coastline must be checked regarding how it is treated in the land-sea mask of the respective GCM. When considering coastal areas, these cells are sometimes considered as a sea cell, with the result of featuring high evaporation and zero or missing runoff quantity. Including such cells in the computation can introduce significant biases and inaccuracies in the computed water budget. So depending on the particular GCM and its land-sea mask, basin-wide quantities require a careful post-processing. The adapted approach fits well for each model, despite their diverse grid resolutions.

It is worth mentioning here that none of the studied climate models implements irrigation or water diversion schemes. Such management schemes are quite relevant for the studied basins, and especially for the Indus basin, to the point that its discharge at the mouth is extremely low. The substantial water diversion from the Indus basin does not allow the direct comparison of the simulated $P - E$ and $R$ with the observed discharge. In order to have at least an approximate comparison, we have estimated the natural discharge from the Indus basin – assuming no diversion within the basin – by summing up the measured discharges into the sea and the amount of runoff diverted for irrigation, industrial and domestic use. The summed up quantity is then transformed into its runoff equivalent. The total water diversion (170 mm yr$^{-1}$) from the Indus basin is around 80 % of its mean surface water availability (i.e. 210 mm yr$^{-1}$) (Laghari et al., 2012). Hence, assuming no diversion within the Indus basin, the net volume of the natural discharge into the sea is roughly around 210 mm yr$^{-1}$ runoff equivalent (or 170 mm yr$^{-1}$ total diverted runoff within the basin + observed 40 mm yr$^{-1}$ runoff equivalent being discharged into the sea). For the other basins, the effects of water management practices are much smaller in relative terms, so that it makes sense to compare the statistics of observed river discharge and of the integrated values of $P - E$ and $R$ as simulated by the models.

## 4 Results

### 4.1 Present climate

A first, important qualitative agreement among all models is that, in agreement with observations, the river basins become wetter as we move eastward, the basic reason being that monsoonal precipitation becomes weaker and weaker the further we move west. Table 3 provides the summary of the long-term means of the simulated basin-integrated hydrological quantities and their inter-annual variability. We discuss here basin-wise the details of our findings about the water balance consistencies and realism of the climate models for the 20th century climate.

#### 4.1.1 Indus Basin

Figure 2a1 shows that most of the models conserve water at a basin scale within a good degree of approximation, regardless of whether their long-term averages of the spatial integral of $P - E$ are realistic or not. However, two models (CNRM and GISS-AOM) have a simulated $P - E$ higher than their respective simulated runoff quantities, implicitly showing that their land surface schemes lose water. On the other hand, two models (IPSL-CM4 and INMCM) show the opposite behavior, with simulated runoff larger than their respective computed $P - E$, showing a gain of water by their land surface schemes. These models feature serious water balance inconsistencies, which may introduce further biases to the regional climate simulations, hence providing unreliable estimates of their projections under warmer climate. In view of these systematic biases/inconsistencies, we suggest that the models should be investigated for relevant diagnostics before their use in further applications.

Figure 2a2 shows the scatterplot of mean annual $P - E$ against its inter-annual variability. Models largely differ with respect to each other for the inter-annual variability of $P - E$, ranging between 30–80 mm yr$^{-1}$. Similarly, models are not consistent with each other for their mean estimates of $P - E$, showing a very large spread of values across 10–350 mm yr$^{-1}$.

As far as models' performance against realism is concerned, it is pertinent to mention here that, as discussed before, the Indus River is highly diverted for irrigation, industrial and domestic use, turning into almost a dry river, whereas GCMs used in the present study do not implement irrigation or water diversion at all. Hence, we compare the average basin-integrated value of $P - E$ simulated by the models with the estimated natural discharge. Figure 2a2 shows the observed as well as the estimated natural discharge of the Indus basin in equivalent runoff units. Most of the models underestimate the mean $P - E$ for the basin when compared with the estimated natural discharge. PCM shows the lowest value of about 10 mm yr$^{-1}$, suggesting almost equal amounts of average precipitation and evaporation, while CNRM model shows the highest value of mean $P - E$ (almost 330 mm yr$^{-1}$). The cluster of six models (CSIRO3.0, IPSL-CM4, HADCM3, INMCM, ECHO-G and CGCM2.3.2) around 70–100 mm yr$^{-1}$ is relatively close to the observed mean annual discharge (i.e. 40 mm yr$^{-1}$) which is of course not relevant as none of these climate models implement irrigation/diversion. Figure 2a3 shows that the main reason of underestimating the estimated natural discharge by these six models is the underestimation (overestimation) of observed precipitation (evaporation). On the other hand, three models (MIROC-HIRES, ECHAM5 and GISS-AOM) modestly agree on $P - E$ with the estimated natural discharge as well as on the observed values of precipitation and evaporation. However, as GISS-AOM features a water balance inconsistency, we do not put much confidence on its performance, which may be a result of different biases cancelling out each other. The so-called ensemble mean is at 150 mm yr$^{-1}$ with only HADGEM1 close to it.

Summarizing the above-mentioned, four models (CNRM, GISS-AOM, IPSL-CM4 and INMCM) do not conserve water for the Indus basin. Models show large differences for the mean $P - E$ and its inter-annual variability. Most of the models underestimate the observed $P - E$ mainly due to the underestimation (overestimation) of observed precipitation (evaporation). A few models (MIROC-HIRES, ECHAM5 and GISS-AOM) agree with the observed precipitation, evaporation and estimated natural runoff.

#### 4.1.2 Ganges Basin

Figure 2b1 shows that INMCM model features unphysical negative estimates of the basin-integrated $P - E$. This is mainly due to the reason that, in the arid and semi-arid areas of the basin, the model suggests a relatively high evaporation as compared to precipitation as well as no runoff (surface and sub-surface runoff) (not shown). This situation influences its basin-integrated $P - E$ estimates, so that a negative value is obtained. This implies non-conservation of water mass with a substantial gain of "ghost" water by the land surface scheme of the model. The same model also substantially underestimates the observed precipitation. On the other hand, IPSL-CM4 has a vanishing value for the basin-integrated $P - E$, while the basin's runoff is positive (see Fig. 2b3). IPSL-CM4 also features the weakest precipitation among the studied models. The rest of the models show good agreement between the computed $P - E$ and the simulated runoff within their associated statistical uncertainties, suggesting that their water balance is closed to a good degree of precision in this basin.

**Table 3.** Long-term means (inter-annual variability) in mm yr$^{-1}$ for 20th, 21st (13 models) and 22nd (10 models) century climates.

| Century | ↓GCM Basin → Basin integrated quantity → | Indus | | | Ganges | | | Brahmaputra | | | Mekong | | |
|---|---|---|---|---|---|---|---|---|---|---|---|---|---|
| | | $P$ | $E$ | $R$ | $P$ | $E$ | $R$ | $P$ | $E$ | $R$ | $P$ | $E$ | $R$ |
| 20th century climate (1961–2000) | CGCM2.3.2 | 317 (76) | 213 (33) | 110 (35) | 508 (86) | 364 (54) | 148 (30) | 1299 (90) | 738 (15) | 570 (86) | 1472 (103) | 866 (18) | 606 (94) |
| | CNRM | 953 (86) | 616 (21) | 288 (53) | 1474 (120) | 927 (34) | 539 (98) | 1221 (131) | 801 (38) | 430 (73) | 1498 (134) | 1075 (52) | 404 (97) |
| | CSIRO3.0 | 301 (56) | 197 (34) | 116 (31) | 691 (122) | 501 (63) | 197 (53) | 1176 (126) | 648 (40) | 546 (103) | 1254 (133) | 803 (53) | 465 (92) |
| | ECHAM5 | 438 (93) | 244 (34) | 199 (51) | 1063 (186) | 622 (60) | 442 (122) | 2319 (269) | 688 (34) | 1640 (270) | 1249 (85) | 978 (59) | 275 (52) |
| | ECHO-G | 316 (61) | 238 (35) | 88 (28) | 617 (94) | 526 (63) | 93 (34) | 694 (76) | 522 (46) | 167 (42) | 1560 (97) | 1120 (53) | 445 (78) |
| | GFDL2.0 | 585 (82) | 301 (39) | 294 (58) | 905 (90) | 477 (50) | 435 (69) | 1380 (203) | 687 (34) | 720 (201) | 1597 (147) | 762 (55) | 863 (132) |
| | HADCM3 | 383 (78) | 288 (37) | 101 (56) | 759 (117) | 519 (54) | 242 (62) | 1485 (130) | 754 (21) | 742 (125) | 1483 (125) | 1077 (34) | 406 (94) |
| | HADGEM1 | 497 (82) | 342 (27) | 171 (52) | 951 (150) | 556 (58) | 430 (98) | 2169 (110) | 782 (15) | 1418 (112) | 1676 (158) | 1063 (29) | 653 (140) |
| | INMCM | 332 (63) | 259 (25) | 120 (31) | 490 (131) | 563 (56) | 82 (38) | 761 (138) | 742 (72) | 116 (35) | 1121 (155) | 903 (76) | 253 (78) |
| | IPSL-CM4 | 429 (74) | 346 (13) | 156 (45) | 395 (106) | 410 (51) | 80 (45) | 693 (105) | 669 (24) | 194 (74) | 1235 (124) | 834 (20) | 448 (106) |
| | MIROC (hires) | 465 (58) | 256 (22) | 211 (28) | 1257 (125) | 703 (36) | 561 (69) | 2188 (167) | 617 (13) | 1579 (170) | 1891 (123) | 1058 (15) | 840 (126) |
| | PCM | 367 (83) | 357 (46) | 12 (30) | 840 (132) | 660 (73) | 160 (47) | 1225 (154) | 723 (50) | 444 (131) | 1148 (79) | 943 (34) | 185 (47) |
| | GISS-AOM | 483 (60) | 238 (17) | 105 (38) | 882 (76) | 571 (31) | 326 (46) | 2571 (136) | 1180 (24) | 1386 (129) | 1700 (59) | 1408 (31) | 321 (22) |
| 21st century climate (2061–2100) | CGCM2.3.2 | 347 (96) | 248 (49) | 103 (51) | 721 (145) | 504 (59) | 219 (59) | 1445 (90) | 804 (20) | 646 (88) | 1530 (114) | 906 (21) | 626 (96) |
| | CNRM | 1008 (91) | 658 (19) | 379 (66) | 1671 (137) | 979 (38) | 712 (124) | 1392 (190) | 914 (87) | 613 (89) | 1555 (177) | 1098 (67) | 434 (115) |
| | CSIRO3.0 | 318 (64) | 217 (35) | 109 (34) | 680 (99) | 499 (55) | 188 (51) | 1179 (133) | 686 (44) | 508 (107) | 1298 (128) | 794 (52) | 515 (98) |
| | ECHAM5 | 375 (78) | 223 (27) | 154 (47) | 1031 (145) | 647 (65) | 386 (86) | 2270 (234) | 727 (39) | 1546 (234) | 1410 (98) | 1036 (55) | 376 (67) |
| | ECHO-G | 364 (87) | 293 (48) | 82 (37) | 867 (99) | 682 (48) | 187 (49) | 812 (82) | 551 (44) | 256 (53) | 1621 (90) | 1112 (46) | 513 (87) |
| | GFDL2.0 | 482 (102) | 280 (44) | 211 (66) | 1001 (178) | 443 (66) | 565 (142) | 2044 (310) | 742 (46) | 1325 (321) | 1624 (166) | 756 (45) | 895 (148) |
| | HADCM3 | 435 (70) | 320 (42) | 120 (58) | 936 (160) | 557 (64) | 379 (104) | 1644 (122) | 768 (22) | 886 (123) | 1550 (122) | 1053 (31) | 501 (100) |
| | HADGEM1 | 512 (64) | 349 (22) | 181 (46) | 1051 (176) | 518 (57) | 563 (106) | 2494 (171) | 789 (18) | 1731 (163) | 1831 (151) | 1010 (34) | 868 (124) |
| | INMCM | 258 (51) | 259 (32) | 64 (16) | 466 (107) | 567 (50) | 83 (32) | 774 (130) | 788 (61) | 113 (40) | 1155 (133) | 930 (49) | 267 (76) |
| | IPSL-CM4 | 412 (68) | 362 (17) | 343 (51) | 288 (88) | 362 (51) | 48 (24) | 515 (119) | 592 (35) | 157 (67) | 1218 (144) | 818 (38) | 459 (130) |
| | MIROC (hires) | 497 (83) | 289 (26) | 210 (44) | 1408 (139) | 718 (37) | 696 (94) | 2937 (345) | 734 (21) | 2210 (345) | 1977 (120) | 1120 (15) | 868 (107) |
| | PCM | 451 (106) | 429 (57) | 21 (31) | 949 (129) | 718 (62) | 204 (61) | 1313 (141) | 740 (42) | 508 (131) | 1189 (86) | 969 (35) | 197 (59) |
| | GISS-AOM | 547 (67) | 267 (20) | 369 (50) | 1068 (107) | 660 (37) | 439 (67) | 2812 (172) | 1283 (25) | 1556 (157) | 1776 (65) | 1443 (36) | 366 (27) |
| 22nd century climate (2161–2200) | CGCM2.3.2 | 354 (73) | 258 (35) | 101 (44) | 685 (144) | 487 (67) | 202 (59) | 1439 (106) | 814 (15) | 630 (96) | 1529 (106) | 910 (20) | 619 (88) |
| | CNRM | 984 (92) | 659 (25) | 394 (70) | 1667 (159) | 978 (34) | 723 (146) | 1544 (144) | 987 (48) | 623 (90) | 1572 (154) | 1101 (58) | 445 (109) |
| | CSIRO3.0 | 351 (77) | 238 (38) | 121 (42) | 694 (113) | 514 (65) | 186 (47) | 1198 (137) | 703 (42) | 509 (107) | 1266 (110) | 779 (40) | 497 (83) |
| | ECHAM5 | 396 (92) | 239 (34) | 160 (58) | 1071 (174) | 681 (66) | 393 (98) | 2244 (222) | 734 (41) | 1513 (214) | 1448 (117) | 1033 (58) | 417 (85) |
| | ECHO-G | 376 (78) | 308 (41) | 79 (35) | 947 (103) | 714 (56) | 235 (62) | 888 (83) | 574 (52) | 312 (57) | 1610 (85) | 1117 (54) | 495 (74) |
| | GFDL2.0 | 453 (79) | 280 (35) | 180 (58) | 999 (199) | 432 (55) | 572 (162) | 2190 (281) | 746 (45) | 1464 (292) | 1613 (171) | 757 (58) | 884 (134) |
| | HADCM3 | 520 (103) | 350 (47) | 175 (74) | 1020 (160) | 562 (49) | 459 (116) | 1771 (161) | 777 (20) | 1000 (165) | 1672 (180) | 1025 (36) | 647 (159) |
| | HADGEM1 | 544 (72) | 371 (26) | 191 (50) | 1102 (189) | 537 (53) | 596 (123) | 2615 (149) | 806 (21) | 1830 (145) | 1706 (169) | 994 (34) | 764 (142) |
| | INMCM | 238 (45) | 249 (24) | 55 (13) | 394 (102) | 533 (43) | 64 (30) | 694 (109) | 749 (56) | 94 (30) | 1111 (143) | 898 (53) | 248 (68) |
| | IPSL-CM4 | 480 (98) | 388 (15) | 329 (60) | 331 (102) | 393 (64) | 62 (35) | 448 (108) | 561 (42) | 109 (47) | 1157 (154) | 799 (49) | 422 (134) |

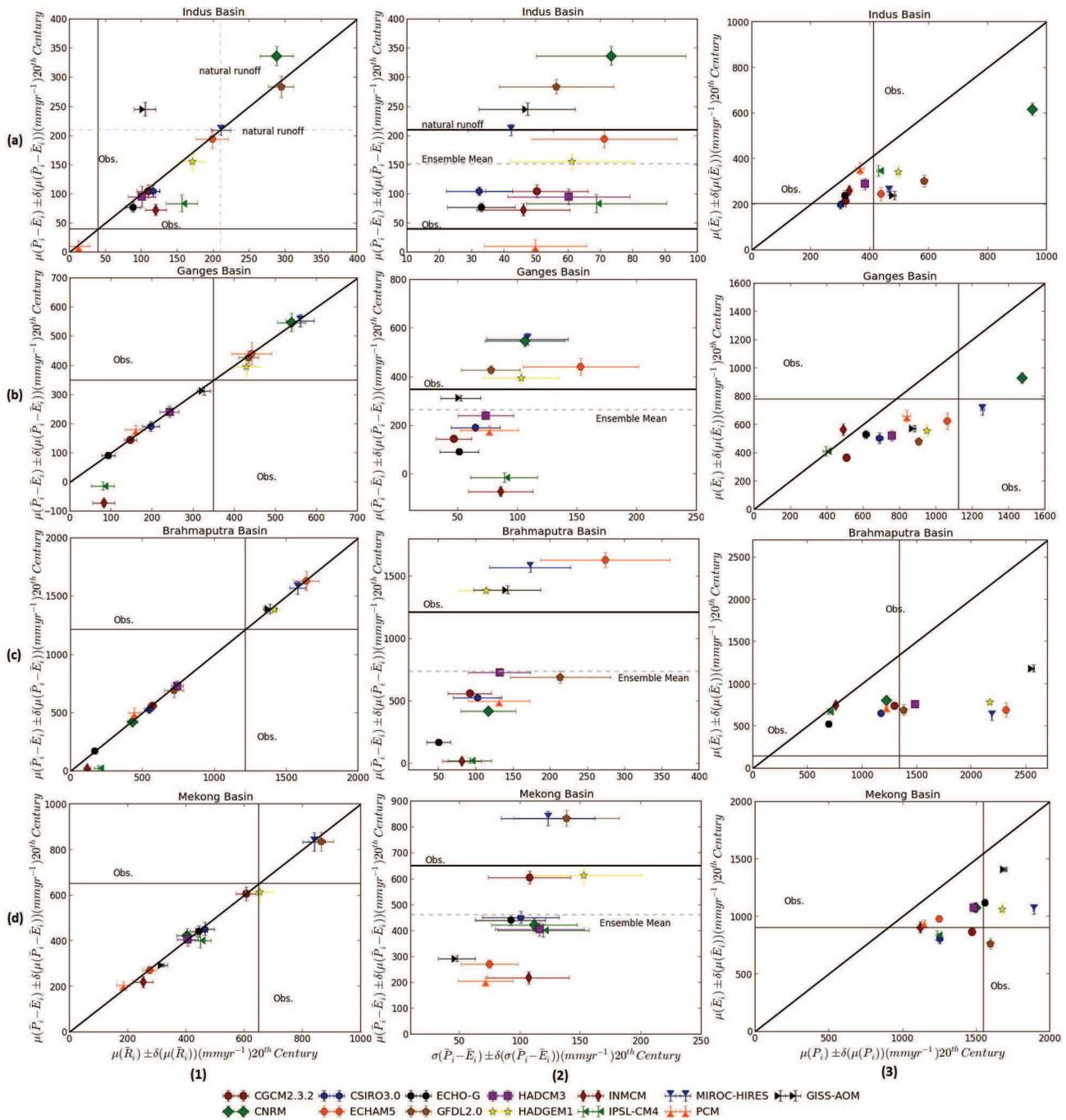

**Fig. 2. (a1)–(d3)** 20th century climate (1961–2000) for: (row a) Indus Basin, (row b) Ganges Basin, (row c) Brahmaputra Basin, (row d) Mekong Basin; (column 1) estimated $P - E$ against simulated total runoff, (column 2) estimated $P - E$ against its inter-annual variability, (column 3) $E$ against $P$. Markers show mean annual simulated basin-integrated quantities, whereas lines show their 95 % confidence intervals. (Note: for the Indus Basin both the observed runoff into the sea (influenced with irrigation and other diversions) and the estimated natural runoff (assuming no diversion within the basin) are shown – details are given in the text.)

Models neither agree well for the mean annual $P - E$ nor for its inter-annual variability (Fig. 2b2). The inter-annual variability of $P - E$ ranges between approximately 50–150 mm yr$^{-1}$. A cluster of six models (CGCM2.3.2, GISS-AOM, HADCM3, CSIRO3.0, PCM and ECHO-G) is centered at 60 mm yr$^{-1}$ and a cluster of five models (CNRM, MIROC-HIRES, HADGEM1, INMCM and IPSL-CM4) is centered at 100 mm yr$^{-1}$. The GCMs outputs for the mean basin-integrated $P - E$ do not agree with each other, showing remarkable spread; however five models (HADCM3, CSIRO3.0, PCM, CGCM2.3.2 and ECHO-G) form a cluster around 200 mm yr$^{-1}$, thus generally below the ensemble mean ($\sim$ 260 mm yr$^{-1}$). The observed mean runoff is around 350 mm yr$^{-1}$ with a modest agreement from four models (GISS-AOM, HADGEM1, ECHAM5 and GFDL2.0). Among these, the latter three underestimate the observed precipitation, i.e. 1125 mm yr$^{-1}$ (CRU, 2012) but slightly overestimate $P - E$ against the observed discharge (i.e. 345 mm yr$^{-1}$), indicating an underestimation of the observed evaporation, whereas GISS-AOM suggests an underestimation of $P - E$, $P$ and $E$.

Figure 2c3 shows that the ratios between precipitation and evaporation are different among the models. All models are in the range of 400–1500 mm yr$^{-1}$ for precipitation against 300–900 mm yr$^{-1}$ for evaporation. Most of the models tend to form a cluster for evaporation around 500 mm r$^{-1}$, showing a good inter-model agreement, however models underestimate the estimated value of observed evaporation.

Summarizing what described above, two models (IPSL-CM4 and INMCM) do not conserve water for the Ganges basin, with a substantial gain of water by their land surface schemes. Models show large differences for the mean $P - E$ but modest agreement for its inter-annual variability. Most of the models underestimate observed $P - E$, which is mainly due to the underestimation of precipitation by these models. A few models (GISS-AOM, HADGEM1, ECHAM5 and GFDL2.0) show modest agreement with the observed runoff but underestimate the precipitation. It is noted that most of the models agree well on the fact that the Ganges is a relatively wetter basin than the Indus Basin.

### 4.1.3 Brahmaputra Basin

Figure 2c1 shows that two models (IPSL-CM4 and INMCM) feature water balance inconsistencies; while the climatology of the integrated value of $P - E$ nearly vanishes, a small but positive runoff is realized. For all the other models, there is a good agreement between the mean simulated runoff and $P - E$ within their associated statistical uncertainties, suggesting that their water balance is closed for the basin. Again, no model is close to the observed runoff of the Brahmaputra Basin.

The inter-annual variability of the $P - E$ shows good inter-model agreement as most of the models cluster around 120 mm yr$^{-1}$, except three models (ECHAM5, MIROC-HIRES and GFDL2.0), which show higher inter-annual variability (Fig. 2c2). However, the overall spread of the inter-annual variability is large (i.e. 50–270 mm yr$^{-1}$). Similarly, models do not agree with each other on the mean value of $P - E$, ranging from almost zero to 1600 mm yr$^{-1}$. Three models (IPSL-CM4, INMCM and ECHO-G) feature the lowest precipitation for the basin. Four models (CGCM2.3.2, CSIRO3.0, PCM and CNRM) form a cluster for $P - E$ around 500 mm yr$^{-1}$ but underestimate it against the mean observed runoff of approximately 1200 mm yr$^{-1}$.

Six models (CGCM2.3.2, CSIRO3.0, GFDL2.0, HADCM3, PCM and CNRM) form a cluster for precipitation around 1250 mm yr$^{-1}$, and nearly agree with the observed precipitation. Six models (CNRM, CGCM2.3.2, CSIRO3.0, PCM, HADCM3 and GFDL2.0) overall show a modest inter-model agreement for the simulated precipitation, evaporation and $P - E$ as well as for the observed precipitation but these models underestimate the observed $P - E$ due to overestimation of observed evaporation. On the other hand, four models (MIROC-HIRES, HADGEM1, GISS-AOM and ECHAM5) overestimate $P - E$ of the basin with the latter one showing the highest value of more than 1600 mm yr$^{-1}$. These models also form a cluster for the precipitation, suggesting highest magnitude of it among the models but a substantial overestimation of it against the observations, i.e. 1350 mm yr$^{-1}$ (CRU, 2012). The HADCM3 and GFDL2.0 models are close to the ensemble mean, which is nearly 740 mm yr$^{-1}$. Figure 2c2 clearly shows large differences among the observed value (i.e. 1200 mm yr$^{-1}$) and the ensemble mean quantity. It is observed that inter-model differences for $P - E$ are generally attributed to the differences in their precipitation regime and overestimation of evaporation. This is also evident from the consistent behavior of evaporation among most of the models (around 600 mm yr$^{-1}$), regardless of the huge differences among their precipitation.

Summarizing the above, two models (IPSL-CM4 and INMCM) do not conserve water for the Brahmaputra basin. Models largely disagree with each other on $P - E$ but show modest agreement for its inter-annual variability. Most of the models underestimate $P - E$ which is mainly due to the underestimation (overestimation) of their simulated precipitation (evaporation). There is an excellent agreement among the models for the Brahmaputra evaporation but our analysis suggests that the models overestimate it.

### 4.1.4 Mekong Basin

Figure 2d1 shows that the water balance of almost all the models is closed within their associated statistical uncertainties for the Mekong basin.

Figure 2d2 shows that most of the models modestly agree on the inter-annual variability of $P - E$ around 100 mm yr$^{-1}$, except five models (GISS-AOM, ECHAM5, PCM, GFDL2.0 and HADGEM1). The spread of inter-annual variability of $P - E$ is about 50–150 mm yr$^{-1}$. Models largely differ in terms of $P - E$, ranging from 200 to 830 mm yr$^{-1}$. Only five models (HADCM3, CNRM, IPSL-CM4, ECHO-G and CSIRO3.0) form a cluster centered at 460 mm yr$^{-1}$, which is also near the ensemble mean. The results show that most of the models underestimate $P - E$ for the Mekong basin against its observed mean runoff of 650 mm yr$^{-1}$. This is mainly associated with the underestimation of precipitation against the observed value of 1550 mm yr$^{-1}$ (CRU, 2012). The underestimation of precipitation may attribute to the fact that the narrow gorge of the northern part of the basin does not fully encompass any coarse resolution GCM grid cell. However, the observed records suggest that the basin mainly receives its precipitation over the southern part of the basin (below 23° N). Such a pattern has also been observed by the GCMs used in our study (not shown). Moreover, such a narrow northern part of the basin covers only the 5 % of the whole basin area. This fact, together with the relative dryness of such a region, makes us fairly confident that the insufficient grid resolution of this area cannot be the cause of the underestimated simulated runoff.

Two models (MIROC-HIRES and GFDL2.0) show the largest values of $P - E$ mainly due to an overestimation of precipitation by the former and an underestimation of evaporation by the latter. There are only two models (HADGEM1 and CGCM2.3.2) which show good agreement with the observed $P - E$ of the basin with slight underestimation. Figure 2d3 shows that this agreement is associated with the realistic evaporation but slight underestimation of precipitation by CGCM2.3.2 but overestimation of the both by HADGEM1 model. The figure also shows that in the case of the Mekong basin, the large spread in $P - E$ is obviously associated with the large inter-model differences for both precipitation (1120–1900 mm yr$^{-1}$) and the evaporation (760–1400 mm yr$^{-1}$).

Summarizing the above-mentioned, almost all the models conserve water for the Mekong basin. Models largely disagree with each other on the mean $P - E$ but show modest agreement for its inter-annual variability. Most of the models underestimate $P - E$, which is mainly due to the underestimation of their simulated precipitation.

## 4.2 Projected changes: 21st and 22nd centuries

Though GCMs' performance in simulating the hydrological quantities is not entirely satisfactory when comparing the outputs to observations, most models feature a consistent water balance. Therefore, it makes sense to investigate how future changes in the hydrological cycle at basin scale are described. In this section we present the results of projected future changes in the hydrological cycle of the four studied basins by CMIP3 climate models. Here, we present the future changes in the hydrological quantities for the latter part of the 21st (13 models) and 22nd (10 models) centuries relative to the latter part of the 20th century by considering the IPCC SRES A1B scenario (Table 3). We observe that going from west to east, GCMs agree qualitatively on the fact that the climate change seems to lead to the wetter and wetter conditions, thus reinforcing the already existing gradient of "wetness".

### 4.2.1 Indus Basin

The inter-annual variability of $P - E$ does not change throughout the 21st and 22nd centuries for the Indus Basin (Fig. 3a), whereas most of the models indicate a decrease in the mean $P - E$ by the 21st and 22nd centuries, with a stronger decrease in the latter century. Interestingly, by looking separately at the changes in $P$ and $E$, one can discover that the response of the various models is less similar than one would guess by looking at $P - E$ alone. The negative change in $P - E$ as simulated by two models (ECHAM5 and GFDL2.0), is due to the stronger decrease in the precipitation rather than in evaporation (Fig. 3a). On the other hand, three models (CGCM2.3.2, CSIRO3.0 and ECHO-G) feature higher positive change in evaporation than in precipitation by the 21st century, and similarly, three models (CNRM, ECHO-G and CGCM2.3.2) by the 22nd century. MIROC-HIRES shows almost the same positive change for both evaporation and precipitation by the 21st century, so that its change in $P - E$ is negligible. Five models (HADCM3, HADGEM1, PCM, CNRM and GISS-AOM) suggest a positive change in $P - E$ by the 21st century, whereas four models (HADCM3, HADGEM1, CSIRO3.0 and IPSL-CM4) suggest a positive change in it by the 22nd century. However, two of these models (CNRM and IPSL-CM4) are largely biased with their water balance inconsistencies.

### 4.2.2 Ganges Basin

Most of the models project an increase in $P - E$ as well as in its inter-annual variability by the end of the 21st and 22nd centuries. However, two models (IPSL-CM4 and IN-MCM) suggest a decrease in the mean $P - E$ and its inter-annual variability for the both 21st and 22nd centuries (Fig. 3b). Such models are affected by the water balance inconsistencies and therefore it is not clear how reliable their projected changes are in these hydrological quantities. ECHAM5 also suggests significant negative changes in $P - E$ for the Ganges basin, which is associated with the negative (positive) changes in the precipitation (evaporation) for the 21st century and mainly with the positive change in the evaporation for the 22nd century. CSIRO3.0 shows a slight negative change in the mean $P - E$ which is associated with the negative change in precipitation for the 21st and 22nd centuries. It also suggests a decrease in the inter-annual variability of $P - E$ for the 21st century but no change in it for the 22nd century. On the other hand, PCM shows an increase in the mean $P - E$ but a decrease in its inter-annual variability for the 21st century.

### 4.2.3 Brahmaputra

Figure 3c shows that most of the models agree for a positive change in $P - E$. Only four models (IPSL-CM4, INMCM, ECHAM5 and CSIRO3.0) suggest a very small negative change in the mean $P - E$ for both 21st and 22nd centuries. As for the Ganges basin, two of these models (IPSL-CM4 and INMCM) do not conserve water and behave quite differently from the rest of the models. Further investigation in terms of changes in the precipitation and evaporation reveals that the negative change in $P - E$ suggested by ECHAM5 is due to the negative (positive) changes in its precipitation (evaporation). Similarly, for CSIRO3.0, this negative change in $P - E$ is mainly due to a positive change in evaporation by the 21st century, and a relatively higher increase in evaporation than in precipitation by the 22nd century. For the Brahmaputra Basin, there is quite good agreement between models for the positive or no change in evaporation.

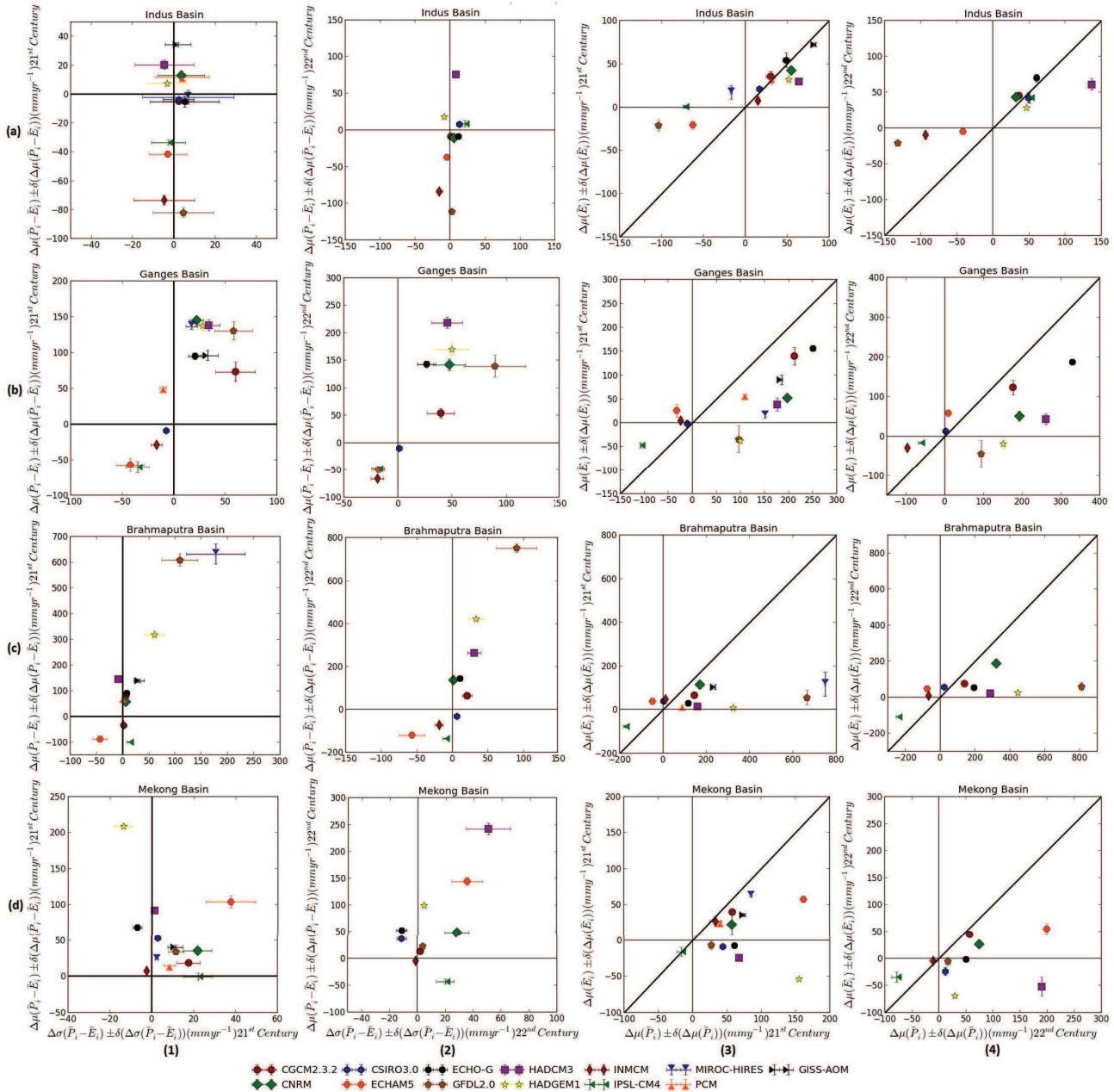

**Fig. 3.** (a1)–(d4) Projected changes for: (row a) Indus Basin (row b) Ganges Basin, (row c) Brahmaputra Basin, (row d) Mekong Basin: (column 1) $P - E$ against its variability for 21st century climate (2061–2100), (column 2) $P - E$ against its variability for 22nd century climate (2161–2200), (column 3) $E$ against $P$ for 21st century climate (2061–2100), (column 4) $E$ against $P$ for 22nd century climate (2161–2200). Markers show mean annual simulated basin-integrated quantities, whereas lines show their 95 % confidence intervals.

As far as the inter-annual variability of $P - E$ is concerned, models agree well on almost no change in it by the 21st century, except ECHAM5 which suggests a decrease, and four models (GISS-AOM, HADGEM1, MIROC-HIRES and GFDL2.0) which suggest an increase. For the 22nd century, most of the models agree either on the slightly positive or no change in the inter-annual variability of $P - E$ where ECHAM5 again suggests a decrease.

### 4.2.4 Mekong Basin

For the 21st century, all the models agree on an increase in the inter-annual variability of $P - E$, except six models (ECHO-G, HADGEM1, INMCM, HADCM3, MIROC-HIRES and CSIRO3.0), whereby the former two suggest a decrease but the latter four suggest almost no change (Fig. 3d). Similarly, for the 22nd century, most of the models suggest an increase in the inter-annual variability of $P - E$, except six models (ECHO-G and CSIRO3.0, INMCM, CGCM2.3.2, GFDL2.0 and HAGEM1), whereby the former two suggest a decrease but the latter four suggest almost no change.

Generally, all GCMs suggest an increase in the basin-integrated mean $P - E$ for the Mekong Basin, even if the agreement of the magnitude of such a change is modest, with HADGEM1 suggesting the highest increase and IPSL-CM4 anticipating almost no change by the 21st century. For the 22nd century, only INMCM shows almost no change for the mean $P - E$ whereas only IPSL-CM4 shows a negative change. Two models (ECHO-G and CSIRO3.0) show an increase in the mean $P - E$ but a decrease in its inter-annual variability.

Figure 3d shows the projected changes in the mean precipitation and the evaporation for the 21st and 22nd centuries. Also in this case, we see that the diversity in the models' response is more pronounced than what could be guessed by looking at the change in $P - E$ alone. For IPSL-CM4, there is a negative projected change in both precipitation and evaporation for the 21st and 22nd centuries. Three models (GFDL2.0, CSIRO3.0, ECHO-G) suggest almost no change while the other two (HADCM3 and HADGEM1) suggest a slight negative change in evaporation but a positive change in precipitation for the 231st and 22nd centuries. For the 22nd century, INMCM suggests almost no change in the precipitation as well as in the evaporation. The rest of the models feature an increase in the both precipitation and the evaporation, the former being stronger than the latter.

## 5 Discussions and conclusions

In this study we have analyzed how the hydrological cycle of four major South and Southeast Asian rivers (Indus, Ganges, Brahmaputra and Mekong) is represented by CMIP3 climate model simulations for the period 1961–2000 (present-day climate) and what future changes are foreseen by these models for the periods 2061–2100 and 2161–2200 under an intermediate warming scenario (IPCC SRES A1B scenario). We have focused on the basin-integrated values of $P$, $E$, $P - E$ and $R$. The focus here has been on assessing how the climate models manage to simulate annual mean conditions over such periods. The inter-model agreement for the simulated hydrological quantities and their relevance to the mean observation is also assessed.

### 5.1 Models performance for the present-day climate

*Conservation of water*: our investigation has shown that most of the CMIP3 climate models are physically consistent in terms of their water balance (Eq. 2) for the studied basins, however a few models present serious deficiencies in conserving water at the basin scale, featuring additional water gain or loss. In particular, four models (CNRM, GISS-AOM, IPSL-CM4 and INMCM) feature water balance inconsistencies for the Indus basin and two models (IPSL-CM4 and INMCM) for the Ganges and Brahmaputra basins, whereas none of the models fails such a consistency check for the Mekong basin. It is generally found that the models having inconsistent water balance over a certain basin also show unrealistic behaviors. We therefore propose using the physical consistency of basic conservation laws as a selection principle (as also proposed by Liepert and Previdi, 2012) and to avoid the use of such models in further analysis or impact assessment studies for the respective basins, as they introduce an unphysical, uncontrolled bias which is impossible to disentangle from the variations induced by anthropogenic forcing.

*Realism*: results found in this study clearly reveal that the models having a consistent behavior in terms of the water conservation at the basin scale also may, firstly, disagree with the observations (runoff and precipitation) and, secondly, largely differ from each other, thus showing a large inter-model spread.

In fact for the Indus basin, three models (MIROC-HIRES, ECHAM5 and GISS-AOM) agree with the estimated natural runoff but these models overestimate the actual precipitation, thus the evaporation from the basin. For the Ganges basin, four models (GFDL2.0, ECHAM5, HADGEM1 and GISS-AOM) show a modest agreement with the observed runoff but underestimate the precipitation and evaporation. For the Brahmaputra basin, no model shows good agreement with the observed runoff. Our analysis also suggests that though there is a good agreement among the models for the Brahmaputra evaporation, models seem to overestimate it. For the Mekong basin, two models (HADGEM1 and CGCM2.3.2) show good agreement with the observed runoff with slight underestimation, which is associated with the realistic evaporation but slightly underestimated precipitation for the CGCM2.3.2 model, whereas an overestimation of both is made by the HADGEM1 model.

Though the overall spreads for the inter-annual variability of $P - E$ are large, most of the models show a modest agreement for all basins except for the Indus basin. For instance, six models (CGCM2.3.2, GISS-AOM, HADCM3, CSIRO3.0, PCM and ECHO-G) form a cluster around 60 mm yr$^{-1}$ whereas five models (CNRM, MIROC-HIRES, HADGEM1, INMCM and IPSL-CM4) form a cluster around 100 mm yr$^{-1}$ for the Ganges basin. For the Brahmaputra basin, most of the models form a cluster around 120 mm yr$^{-1}$ whereas for the Mekong basins it is around 100 mm yr$^{-1}$.

The reason why CMIP3 models fail in simulating the mean $P - E$ over the historical period 1961–2000 is mainly associated with the underestimation of precipitation. The remarkable uncertainty shown by the CMIP3 climate models in simulating the hydrological cycle of the studied basins is unavoidably linked to the deficiencies of these models in simulating realistically the summer monsoon (Turner and Annamalai, 2012). This misrepresentation of the precipitation regime by CMIP3 models can partly be associated with their spatial biases of the Indian summer monsoon precipitation maxima which is shifted towards equator at 12° N (Lin et al., 2008). It is evident from the findings of Boos and Hurley (2013), linking an inaccurate representation of orography to a bias in the thermodynamic structure of the summer monsoon as represented by CMIP3 and CMIP5 models, which results in ensemble-mean negative precipitation anomalies over the Indian region. This is partially consistent with our results since, as explained, most of models tend to underestimate rainfall over the study basins. On the other hand, the large inter-annual variability found in $P - E$ is mainly associated with the variability in the monsoon system, which is also influenced by its interplay with the mid-latitude circulation over the study region, as the westerly troughs can penetrate deeper and suppress the monsoon thermal contrast by the cold air advection over the monsoon dominated region which weakens the monsoon strength (Kripalani et al., 1997; Zickfeld et al., 2005). This interaction causes the monsoon onset delays and also the breaks. The resultant variability in the $P - E$ brings severe implications of the extreme wet and droughts conditions in the region. However, this interaction is not realistically represented by the models due to an inaccurate representation of topography, which is further constrained by their structural characteristics (e.g. resolution). These structural differences among models are possibly one of the major causes of the inter-model spread of the overall precipitation regime over the region.

We have found that the distribution of GCMs' results for the various considered hydro-climatic diagnostics do not look like a well-behaved uni-modal distribution, with data accumulating around a well-defined value. Instead, there is a large degree of inter-model variability and sparse clustering of models' outputs. This is evident, for example, when we look at the mean $P - E$ for all river basins and particularly in case of the Brahmaputra and Mekong. In these conditions, we are of the view that considering the ensemble means of the relevant quantities for verification procedure, neither bears any robust statistical value nor any practical significance, and that the estimates relying on particularly the ensemble quantities can be quite misleading. Furthermore, it is crucial to report the results of the individual models in order to provide information of interest for those who work on statistical or dynamical downscaling. In Table 4 we report the value of a simple indicator, suggesting how peaked is the distribution of GCMs' outputs (for 20th and 21st century only). Such an indicator is constructed as the ratio between the ensemble spread, defined by the maximum minus the minimum (we loosely follow the suggestion by Judd et al., 2007), divided by the ensemble mean. As we see, for most hydroclimatic variables, the value is larger than unity, the only exception being the basin-integrated evaporation and precipitation of the – very wet – Mekong basin. Therefore, it is apparent that the ensemble means tell only a very limited part of the story. Note that in all cases, the indicator is much higher for the quantity $P - E$ than for either $P$ or $E$ (signal to noise ratio is lower for $P - E$), which also suggests that inter-comparing models on precipitation only, e.g. leads to underestimating the inter-model uncertainties in the hydrological cycle.

**Table 4.** Ratio between the ensemble spread and the ensemble mean for the 40 yr climatological averages of selected basin-integrated quantities.

| Period | Basin integrated quantity | Indus | Ganges | Brahmaputra | Mekong |
|---|---|---|---|---|---|
| 20th century climate (1961–2000) | $P$ | 1.4 | 1.3 | 1.3 | 0.5 |
| | $E$ | 1.4 | 1.0 | 0.9 | 0.7 |
| | $P - E$ | 2.2 | 2.4 | 2.2 | 1.4 |
| | $R$ | 1.9 | 1.7 | 2.0 | 1.4 |
| 21st century climate (2061–2100) | $P$ | 1.6 | 1.5 | 1.5 | 0.5 |
| | $E$ | 1.4 | 0.9 | 1.0 | 0.7 |
| | $P - E$ | 2.5 | 2.6 | 2.4 | 1.3 |
| | $R$ | 2.0 | 2.3 | 1.9 | 1.3 |

## 5.2 Projections for 21st and 22nd century climates

The analysis of the 21st and 22nd century, based on the CMIP3 simulations, shows a large spread of simulated hydrological quantities for all the four river basins, which prevents precise quantitative analysis. However, some unequivocal, clear trends emerge from the models' inter-comparison analysis.

First, CMIP3 models generally show an increase in the precipitation and in the simulated runoff over the Ganges, Brahmaputra and Mekong basins under the SRESA1B scenario throughout the 21st and 22nd centuries (Fig. 3). This is generally in agreement with the present knowledge of the effects of increased $CO_2$ levels in the South Asian summer monsoon (Cherchi et al., 2011; Turner and Annamalai, 2012); although a detailed understanding of how the monsoon will change under warmer climate still poses a great challenge to the climate science community. The monsoonal precipitation increase is associated with an increase in the thermal contrast between the Indian continent and the Indian Ocean as well as with the increase in the moisture content of the atmosphere even though the monsoon wind circulation should weaken in the future warmer climate (May, 2002). Such increases in precipitations, however, might not necessarily correspond to the increases in the runoff or the water budget because of the anticipated higher evapotranspiration under warmer climate. Our study shows that most of the models predicting an increase in precipitation, predict an increase in evaporation too, although of a minor magnitude. Furthermore, the increase in $P - E$ is generally associated with an increase in the precipitation of the CMIP3 models.

According to CMIP3 projections, the Ganges and Mekong basins will experience an increased variability in their $P - E$, thus indicating a possible increase in the frequency of extreme low-frequency dry and wet fluctuations. Understanding the future trends and the dynamical mechanisms involved in the extreme events associated with anomalous monsoon circulations will require more research since, in the light of recent extreme events in the area (e.g. the 2010 Pakistan flood) it is of high societal and political value. Going more into each specific case, no considerable change is found in the inter-annual variability of $P - E$ for the Indus and Brahmaputra basins (range $\pm 20$ mm yr$^{-1}$ for the 21st century and similarly for the 22nd century), with a robust agreement between most of the models.

A different scenario is found for the Indus basin, for which most of the CMIP3 models predict a decrease in the mean $P - E$ (and so in the simulated runoff), but no change in its inter-annual variability (Fig. 3a), thus configuring a remarkably different situation with respect to the other three river basins. Such contradictory results are due to the fact that its $P - E$ is determined by a more complex atmospheric circulation, determined by both mid-latitude cyclones and summer monsoon. Almost half of the Indus basin precipitations, in fact, come from the winter snowfalls over the large HKH mountains due to the extra tropical cyclones originating over the Caspian and the Mediterranean sea at the easternmost extremity of the Atlantic and Mediterranean storm tracks (Hodges et al., 2003; Bengtsson et al., 2007). The remaining part of the Indus basin precipitation is due to the summer monsoon system. The Indus basin area therefore, located at the border of these two main large-scale circulations, is indeed a challenge for the climate models to simulate its hydroclimatology realistically. Such complex meteorology does not allow us to say which of the two weather systems will be responsible for the increased dry conditions. In order to understand it, it is worthwhile to decompose the analysis further at the inter-seasonal scale to see how the hydrological cycle is represented at a seasonal timescale and which factors are responsible for the future changes in the relevant hydrological quantities. This is the subject of a companion paper. Furthermore, an important role in future climate will be played by variations in the snowfall over the HKH glaciers, which are mainly fed by winter mid-latitude cyclones, and by the effect of a warmer climate on such glaciers. Particularly in the present-day conditions, snowmelt and rainfall fluctuations are compensated by the glacier melt. But at the moment, it is not clear how this will change in the future. Therefore, it is crucial to analyze/project such behavior correctly in order to understand changes in the Indus basin hydrology. However, assessing the compensation of monsoon rainfall to the glacier melt is challenging. According to our present knowledge about the changes in the monsoon precipitation by the end of the 21st and 22nd centuries, future response of the glaciers to the runoff is also not very clear, especially for the Indus Basin, which has a large portion of its discharge dependent on the glacier melt in addition to the snowmelt. A modeling study suggests a decrease in the Indus basin runoff for the late spring and summer seasons after the period of rapid glacier melt (Immerzeel et al., 2010), which actually depends crucially on how much monsoon precipitation compensates the melt runoff under the warmer climate.

Similar to the Indus, no change is projected in the inter-annual variability of $P - E$ for the Brahmaputra basin by the 21st century, and only a slight increase by the 22nd century. This may be due to the fact that the basin receives almost 10 % of its mean precipitation under the western disturbances (mainly in the form of solid precipitations over the Himalayan Mountains). Under such a scenario, changes in the basin mean $P - E$ and its inter-annual variability is strongly related to the changes in both the monsoonal precipitation over the Indian region (Cherchi et al., 2011; Turner and Annamalai, 2012; Stowasser et al., 2007) as well as in the western disturbances.

Finally, it is worth considering the significance of the ensemble mean in changed climate conditions and the relevance of inter-model uncertainties. One must note – see Table 3 – that when considering the 21st century projections (not the projected differences), the ratio between the ensemble spread and the ensemble mean increases with respect to the 20th century data for $P$, $P - E$ and $R$ for the Indus, Ganges and Brahmaputra (except $R$ in the last river basin), suggesting increasing inter-model uncertainties for these hydro-climatological quantities, while such uncertainties decrease when considering evaporation, probably because it is better constrained by the overall pattern of increases of the surface temperature. Finally, in the case of the Mekong basin the inter-model uncertainties are much smaller than for the other rivers and smaller than in the 20th century data, thus suggesting better model agreement in warmer conditions. This provides further caveats for using ensemble averaged quantities in such complex dynamical contexts.

*Acknowledgements.* The authors thank Dieter Gerten, the editor, and the anonymous reviewers for their comments that improved the manuscript. The authors acknowledge the modeling groups, the Program for Climate Model Diagnosis and Inter-comparison (PCMDI) and the WCRP's Working Group on Coupled Modeling (WGCM) for their roles in making available the WCRP CMIP3 multi-model dataset. Support of this dataset is provided by the Office of Science, US Department of Energy. SH acknowledges the support of BMBF, Germany's Bundle Project CLASH/Climate variability and landscape dynamics in Southeast-Tibet and the eastern Himalaya during the Late Holocene reconstructed from tree rings, soils and climate modeling. V. Lucarini and S. Pascale acknowledge the support of the ERC Starting Investigator grant NAMASTE/Thermodynamics of the Climate System (Grant No. 257106). The authors also acknowledge the support received from CliSAP/Cluster of excellence in the Integrated Climate System Analysis and Prediction. The authors also acknowledge Sabine Ehrenreich for her contribution in editing and improving the manuscript.

Edited by: D. Gerten